\PassOptionsToPackage{nomessages}{fp} 
\documentclass[final,journal,letterpaper,twocolumn,11pt]{IEEEtran}
\usepackage{cite}

\ifCLASSINFOpdf
  \usepackage[pdftex]{graphicx}
  \graphicspath{{./images/}}
\else
  \usepackage[dvips]{graphicx}
  \graphicspath{{./images/}}
\fi
\usepackage[cmex10]{amsmath}
\ifCLASSOPTIONcompsoc
  \usepackage[caption=false,font=normalsize,labelfont=sf,textfont=sf]{subfig}
\else
  \usepackage[caption=false,font=footnotesize]{subfig}
\fi
\usepackage{fixltx2e}
\usepackage[utf8]{inputenc}
\usepackage{amsfonts}
\usepackage{paralist}
\usepackage[shortcuts,nonumberlist,acronym]{glossaries}
\usepackage{pgfplotstable}
\usepackage{booktabs}
\usepackage{todonotes}
\makeglossaries

\def\x{{\mathbf x}}
\DeclareMathOperator*{\argmax}{arg\,max}
\DeclareMathOperator*{\argmin}{arg\,min}

\pgfplotstableset{col sep=comma}

\usepackage{ifthen}
\let\oldcite=\cite
\renewcommand\cite[1]{\ifthenelse{\equal{#1}{citation_needed}}{\ensuremath{^\texttt{[citation~needed]}}}{\oldcite{#1}}}
\begin{document}

\newacronym{pca}{PCA}{principal component analysis}
\newacronym[longplural={principal components}]{princomp}{PC}{principal component}
\newacronym{lda}{LDA}{linear discriminant analysis}
\newacronym{qda}{QDA}{quadratic discriminant analysis}
\newacronym{pda}{PDA}{penalised discriminant analysis}
\newacronym[longplural={support vector machines}]{svm}{SVM}{support vector machine}
\newacronym{ecg}{ECG}{electrocardiogram}
\newacronym{emd}{EMD}{empirical mode decomposition}
\newacronym[longplural={intrinsic mode functions}]{imf}{IMF}{intrinsic mode function}
\newacronym[longplural={automated external defibrillators}]{aed}{AED}{automated external defibrillator}
\newacronym[longplural={automatic implantable cardiovertor defibrillators}]{icd}{AICD}{automatic implantable cardiovertor defibrillator}
\newacronym{psa}{PSA}{phase space algorithm}
\newacronym{ops}{OPS}{original phase space algorithm}
\newacronym{psm}{PSM}{phase space modified}
\newacronym{rbf}{RBF}{radial basis function}
\newacronym{max-likeli}{ML}{maximum likelihood}
\newacronym{loo}{LOO}{leave one out}
\newacronym{aha}{AHA}{American Heart Association}
\newacronym{snr}{SNR}{signal to noise ratio}
\newacronym{srhythm}{SR}{sinus rhythm}
\newacronym{v-fib}{VF}{ventricular fibrillation}
\newacronym{v-tach}{VT}{ventricular tachycardia}
\newacronym{tdp}{TDP}{torsade de pointes}
\newacronym{a-fib}{AF}{atrial fibrillation}
\newacronym{cpr}{CPR}{cardiopulmonary resuscitation}
\newacronym{torsades}{TdP}{Torsades de Pointes}
\newacronym{lqts}{LQTS}{long Q-T syndrome}
\newacronym{world_health_org}{WHO}{World Health Organisation}
\newacronym{cvd}{CVD}{cardiovascular disease}
\newacronym{edb}{EDB}{European ST-T Database}
\newacronym{vfdb}{VFDB}{MIT-BIH Malignant Ventricular Arrhythmia Database}
\newacronym{cudb}{CUDB}{Creighton University Ventricular Tachyarrhythmia Database}
\newacronym{mitdb}{MITDB}{MIT-BIH Arrhythmia Database}
\newacronym{ahadb}{AHA}{American Heart Association Database}

%
\title{Classification of Human Ventricular Arrhythmia in High Dimensional Representation Spaces}
%
%
%

\author{Yaqub~Alwan,
        Zoran~Cvetkovi\'c,~\IEEEmembership{Senior~Member,~IEEE,}
        Michael~J.~Curtis
}%

%
%

\markboth{}%
{Alwan \MakeLowercase{\textit{et al.}}: Classification of Human Ventricular Arrhythmia in High Dimensional Representation Spaces}%
%



\maketitle

\begin{abstract}

We studied classification of human \acrshortpl{ecg} labelled as normal \acl{srhythm},
\acl{v-fib} and \acl{v-tach} 
by means of \acrlongpl{svm} in different representation spaces, using different
observation lengths. \acs{ecg} waveform segments of duration 0.5-4~s, their Fourier
magnitude spectra, and lower dimensional projections of Fourier magnitude spectra
were used for classification. All considered representations were of much higher
dimension than in published studies. Classification accuracy improved with segment
duration up to 2~s, with 4~s providing little improvement.We found that it is
possible to discriminate between \acl{v-tach} and \acl{v-fib} by the present
approach with much shorter runs of \acs{ecg} (2~s, minimum 86\% sensitivity per class)
than previously imagined. Ensembles of classifiers acting on 1~s segments taken
over 5~s observation windows gave best results, with sensitivities of detection
for all classes exceeding 93\%. 

\end{abstract}


%
\IEEEpeerreviewmaketitle

\section{Introduction}
\label{sec:intro}
%
%
%
%
\IEEEPARstart{A}{ccording} to the World Health Organisation data, \acl{cvd} is the leading cause of death
in middle and high income countries, and among the top ten causes of death in low income countries
\cite{who_top_10_causes_death}. Development of effective drug treatments that may prevent cardiac arrhythmia
is therefore a high-priority challenge for modern pharmacology. For the development of such treatments it is
crucial to have a clear understanding of what distinguishes different forms of arrhythmia, and based on that,
establish their precise definitions.
However, it is evident that although unequivocal \gls{v-fib}, sustained and lethal, is incontestable in \gls{ecg}
recordings, clinicians differ fundamentally about the diagnosis and appellation of transient polymorphic
ventricular tachyarrhythmias, with experts in a landmark report unable to agree on whether \gls{v-fib},
polymorphic \gls{v-tach} or \gls{tdp} best described a range of human tachyarrhythmias in a blinded test of
\gls{ecg} records~\cite{self_terminating_va_dilemma}.
This attests to limitations in definitions/appellation. Given that mechanisms of these tachyarrhythmias may
differ~\cite{ph2_va_infarction} and responses to drugs may vary from benefit to proarrhythmia, depending on
the type~\cite{effects_atp_potassium_activation}, errors in diagnosis due to unequivocal appellations are potentially
hazardous. 
To allow preclinical research to be translatable, a definition was recently
proposed to discriminate between \gls{v-fib}, including brief and transient \gls{v-fib},
and other polymorphic ventricular
tachyarrhythmias~\cite{lambeth_conventions_13}. This definition is not, however, readily transformed into an
algorithm for automatic rhythm detection.
Therefore, in the present study we develop novel algorithms for ventricular tachyarrhythmia classification
and use them to assess 
 whether it is possible to improve precision and accuracy of discrimination
between \gls{v-fib} and \gls{v-tach}.

There have been many studies into the topic of differentiating \gls{srhythm} from \gls{v-fib},
however fewer studies attempt to differentiate \gls{v-tach} from \gls{v-fib}. From a therapeutic
point of view being able to differentiate between \gls{v-fib} and  \gls{v-tach} is very important
since they respond to interventions differently and \gls{v-fib} is often lethal, while \gls{v-tach} is often not.
For \gls{v-fib} and \gls{v-tach} detection, Thakor {\sl et al.}
developed an algorithm which first applies a hard threshold to transform  an \gls{ecg} segment into a binary
sequence, and then performs a sequential hypothesis test on the average number of zero crossings until a
decision is made~\cite{arrhythmia_detection_sequential_hypothesis_tci}. In~\cite{wavelet_features_for_characterising_arrhythmia},
the authors proposed an algorithm which uses the energy distribution information in a wavelet transform domain to
differentiate between \gls{v-fib}, \gls{v-tach}, and a \gls{v-tach}-\gls{v-fib} class which contains realisations
that are difficult to categorise as either \gls{v-tach} or \gls{v-fib}. On the other hand,
an estimate of the area occupied in the bispectral
representation is proposed for classification between \gls{v-fib}, \gls{v-tach}, \gls{srhythm} and \acl{a-fib} in
\cite{cardiac_arrhythmia_classification_using_high_order_spectral_techniques}. The standard deviations of the peak amplitudes and
peak distances in a cross correlation between a window of \gls{ecg} and a short window immediately prior is used to classify
\gls{v-tach} from \gls{v-fib} in~\cite{detection_arrhythmia_using_crosscorrelation}. An algorithm for \gls{v-fib} detection based
on \gls{emd}, which creates a lower dimensional representation of \gls{ecg} waveforms based on the energy in the first
\gls{emd} component and its spectral entropy, was proposed in~\cite{vf_detection_emd}. Counting the time between \gls{ecg} turning
points was proposed~\cite{discrimination_vt_vf_for_icd} to differentiate fast-\gls{v-tach}, slow-\gls{v-tach} and
\gls{v-fib} for use in \glspl{icd}. A comparison of ten methods for differentiating between non-\gls{v-fib} and \gls{v-fib},
for use in an \gls{aed}, was presented in~\cite{reliability_vf_detection_for_aed}. The same authors subsequently proposed a phase
space method~\cite{detecting_vf_time_delay_methods} and demonstrate that it outperforms all methods studied in
\cite{reliability_vf_detection_for_aed} in its capability to discriminate \gls{v-fib} from non-\gls{v-fib} segments. While
all these previous works have been yielding gradual improvements, the problem of differentiating between \gls{srhythm},
\gls{v-tach} and \gls{v-fib} is still not solved sufficiently accurately, the main difficulty being in differentiating
between \gls{v-tach} and \gls{v-fib}~\cite{eusipcopaper}. 
Some of these studies even resort to creating a category specifically for the examples
which are difficult to distinguish~\cite{wavelet_features_for_characterising_arrhythmia,discrimination_vt_vf_for_icd}. Often, studies are conducted
with limited or preselected data. In addition, all existing classification algorithms use heuristics to derive some
low-dimensional representations of \gls{ecg} signals and decision parameters, rather than provide a data-driven insight into what distinguishes one
arrhythmia from another.

There is therefore a need for systematic investigation of classification of cardiac arrhythmias using representations which involve
minimal information reduction, or data driven dimension reduction and classification using well established and understood
statistical methods. We focus on discrimination between \gls{srhythm}, \gls{v-tach}, and \gls{v-fib},
using as few heuristics as possible. In a preliminary study we considered classification in the
domain of the \gls{ecg} signal, and its Fourier magnitude spectra, using \gls{lda}, \gls{pda} and \gls{qda}, combined with
\gls{pca} for data driven dimension reduction~\cite{eusipcopaper}. 
We concluded that there is benefit in
considering non-linear class boundaries, but that \gls{qda} may not
be the  most suitable type of non-linearity, and that  the training data was insufficient to obtain good generalization.
In an attempt to address these issues,
in the present study we considered classification with the same representation spaces using
\glspl{svm}~\cite{svm_1995}.

One issue of interest is the selection of observation length which is most suitable for discrimination,
whilst simultaneously trying to minimise detection time.
In order to avoid introducing low-dimensional heuristic
features, classification is performed directly in the space of \gls{ecg} waveforms, and their
Fourier magnitude spectra. Further, lower-dimensional representations obtained by projecting magnitude
spectra onto respective \gls{princomp} subspaces are also considered, in order to facilitate learning of non-linear
boundaries using relatively limited training data. Even with this dimension reduction, the lowest
representation dimension considered in this study is \(10\), which is significantly higher dimension than features
vectors used in prior art. Experimental results reported in Section~\ref{sec:experiments} demonstrate the benefits
of considering such higher-dimensional representations for classification purposes.

The paper is organised as follows. In Section \ref{sec:problem} we review a reference prior work, provide rationale for shorter
observation windows, and specify dimension-reduction methods used. Section \ref{sec:svc} provides details of
the \gls{svm} classification framework used. Experimental procedure and results are reported in Section \ref{sec:experiments}. Section
\ref{sec:conclusion} summarises the paper and draws conclusions.

\section{General Considerations}
\label{sec:problem}

Given an \gls{ecg} segment 
\[\x=\{x[n], n_1\le n\le n_2\}~,\] 
it is desired to be able to label it as being \gls{srhythm}, \gls{v-tach} or \gls{v-fib}. For comparison, in this paper
the phase space feature representation proposed in~\cite{detecting_vf_time_delay_methods} is considered, since the authors conducted
a thorough investigation and demonstrated its superior performance in terms of accuracy and numerical complexity compared to previously
published algorithms.

\subsection{Reference prior art}
\label{subsec:priorart}

The \gls{psa}~\cite{detecting_vf_time_delay_methods} aims at diagnosing whether a defibrillation shock should be delivered,
which amounts to classifying \gls{ecg} segments as \gls{v-fib} or non-\gls{v-fib}. The phase representation is formed by taking
discretised values of samples of \(\x\) as pairs $(x_1[n],x_2[n])$ in \(\mathbb{R}^2\), where
\begin{eqnarray*} 
x_1[n] & =  & x[n] \\
x_2[n] & =  & x[n-k]
\end{eqnarray*}
while \(k\) is selected to correspond to \(0.5\)~s\footnote{These parameters are all selected
by the original work,~\cite{detecting_vf_time_delay_methods}}. The discretisation step is chosen such that the complete range of
\(x_1[n]\) and \(x_2[n]\) each take up to \(40\)\footnotemark[\value{footnote}] unique values. The number \(N(\x)\) of visited
boxes in this phase space is then found, and finally, the ratio
\[\eta(\x)=\frac{N(\x)}{N_{\rm max}}\]
between the number of visited boxes and the total number of boxes \(N_{\rm max}=40 \times 40\) is compared to an empirically determined
threshold \(\eta_{\rm thresh} = 0.15\)\footnotemark[\value{footnote}]. If this threshold is exceeded, \gls{v-fib} is decided, otherwise
non-\gls{v-fib} is decided. The representation space formed by \(N(\x)\) (one-dimensional space) is referred to as \gls{psa}. 
We introduce a modification where the phase space is formed by
pairs 
\begin{eqnarray*}
x_1[n] & = & x[n]\\
x_2[n] & = & x[n]-x[n-1]
\end{eqnarray*}
again discretised so that each take up to \(40\) unique values. This corresponds to the standard notion of a phase space,
and it may be more robust to variations in heart rate. This modification is referred to as \gls{psm}. In Section
\ref{sec:experiments} results are reported for both versions of the phase space representation in comparison to
other representations introduced here.

\subsection{Observation length}
\label{subsec:discrim}

One of the important issues that needs to be investigated is the appropriate observation length for reliable
classification of cardiac arrhythmias. Longer windows contain more  information, and that  should  lead to
higher classification accuracy, up to a point beyond which increasing observation length does not provide any
additional discriminating information. This plateau can be expected to be followed by a decline in classification
accuracy when windows are  long enough to include transitions between \gls{ecg} classes. For fast and reliable
diagnosis it is of interest to simultaneously minimise observation length and maximise classification accuracy.
However, a systematic investigation of this issue is lacking in prior studies.

The phase space method in its original formulation~\cite{detecting_vf_time_delay_methods} counts the number of
visited boxes over \(8\)~s segments of \gls{ecg} signals. Several other studies also form their respective feature
vectors using \(8\)~s observations~\cite{reliability_vf_detection_for_aed},
\(6\)~s observations~\cite{detection_arrhythmia_using_crosscorrelation}, or over  \(5\)~s observations
\cite{arrhythmia_detection_sequential_hypothesis_tci}.
Physiological considerations, on the other hand, suggest that shorter \gls{ecg} segments should suffice. According
to~\cite{lambeth_conventions_88}, \gls{v-tach} is considered to occur if \(4\) or more consecutive QRS complexes
precede their corresponding P-wave, independent of the rate. Assuming a heart rate of \(60\) beats per minute or
higher in humans, a \(2\)~s window should be sufficient to capture \(2\) normal beats. Thus, at least \(3\)
premature QRS complexes would occur in the same \(2\)~s interval, and since \gls{v-tach} is usually accompanied
by an increased heart rate, quite often \(2\)~s should also be sufficient to capture \(4\) or more premature QRS
complexes. For \gls{v-fib}, QRS complexes are no longer discernible~\cite{lambeth_conventions_88}, which suggests
the rate of cardiac deflections (not heart rate, since the notion is not applicable) is even higher than that of
\gls{v-tach}. Therefore, a \(2\)~s window should be sufficient for capturing the disorder of \gls{v-fib} also.
Hence, this study will be centred around \(2\)~s windows, while \(0.5\)~s, \(1\)~s, and \(4\)~s windows will be
also considered in order to assess effects of observation length on the classification accuracy and draw some
conclusions on its optimal value.  

\subsection{Dimension reduction}

In order to avoid any bias due to preconceptions on what features are relevant for arrhythmia classification,
classification directly in the domain of \gls{ecg} waveforms is considered as a starting point. Further,
Fourier magnitude spectra of \gls{ecg} segments are considered as they provide representations invariant to
time shifts while preserving most of the information contained in the original waveforms. At \(100\)~Hz
sampling, which is close to the minimal sampling frequency which does not cause apparent distortions of
\gls{ecg} signals, \(4\)~s observation length results in \(400\)-dimensional feature space
(or \(200\)-dimensional space if magnitude spectra are considered), which can make statistical inference
challenging. This is dealt with by employing data-driven dimension reduction based on \gls{pca}.
We performed \gls{pca} on magnitude spectra of each individual class, and formed sets of basis vectors by the union of
top \(N\) \acrlongpl{princomp} from each class. Since at least 5~\acrlongpl{princomp} are required to capture at least
\(60\%\) of the energy for each class, this is the lower limit considered. For the two class case, this forms a space of dimension \(10\),
and \(15\) in the three-class case. The upper limit on the number of principal directions
considered is \(15\), since already then the dimension of the complete representation for three-class tasks becomes
as high a \(45\), which exceeds the dimension of \(0.5\)~s magnitude spectra. Since class labels are
utilised, this step is performed as a supervised step in experiments reported in \ref{sec:experiments},
only having access to the training set for the purposes of learning the basis vectors.

\section{Classification using \acrlongpl{svm}}
\label{sec:svc}

Given a set of training data \(\left(\x_1,\dotsc,\x_p\right)\) with corresponding class labels
\(\left(y_1,\dotsc,y_p\right),~y_i \in \{ +1,-1 \}\), an \gls{svm} attempts to find a decision
surface which jointly maximizes the margin between the two classes and minimizes the misclassification
error on the training set. When the classes are linearly separable, these surfaces are linear and have the form
\begin{equation}
\label{equ:linear_svm_surface}
f(\mathbf{x}) = \sum_{i}a_{i}y_{i} \langle \x,\x_i \rangle + b=0,~a_{i} \in \mathbb{R}^+
\end{equation}
where \(\langle \cdot, \cdot \rangle\) is the inner product in \(\mathbb{R}^n\), while the Lagrange multipliers \(a_i\)
and the bias \(b\) are optimized  by the training algorithm. Non-linear separators between two classes are
created by means of non-linear kernel functions \(K(\cdot,\cdot)\). These functions compute inner products in higher
dimensional spaces, without explicitly performing the mapping, where the data could potentially be linearly separable.
Analogously to the linearly separable case, the decision surface is constructed according to
\begin{equation}
\label{equ:kernel_svm_surface}
f(\mathbf{x}) = \sum_{i}a_{i}y_{i} K( \mathbf{x},\mathbf{x}_i ) + b = 0
\end{equation}
and the class label of a test vector \(\x\) is predicted to be the sign of the score function evaluated at \(\x\), {\sl i.e.}
\[C(\x) = \mathop{\rm sgn}(f(\x))~.\]

One commonly used kernel function is the polynomial kernel, of the form
\begin{equation}
\label{equ:poly_kernel}
K_{p}(x,y) = (1 + c \langle x,y \rangle)^d, d \in \mathbb{N}
\end{equation}
where \(c\) and \(d\) are optimised using a grid search. Another kernel function is the
\gls{rbf} kernel given by
\begin{equation}
\label{equ:rbf_kernel}
K_r(x,y) = e^{-\gamma\lvert\lvert x - y \rvert\rvert^2}, \gamma \in \mathbb{R}^+
\end{equation}
where \(\gamma\) is also optimised using a grid search.

\subsection{Optimising \gls{svm} parameters}
\label{subsec:svm_parameters}

Unlike with \gls{lda} and \gls{qda}, there are \gls{svm} parameters that need to be chosen optimally for best results.
The idea is to avoid selecting parameters too finely tuned to the training data, while still giving good
classification performance on unseen examples (test data). For \glspl{svm} using either the linear,
polynomial, or \gls{rbf} kernels, these parameters are:

\(\mathcal{C}\): Trade-off between margin width and misclassified examples from the training data

\(\gamma\): \gls{rbf} kernel parameter which controls the width of the Gaussian function. Small
values of \(\gamma\) lead to increasing flexibility of the decision boundary, while large values
of \(\gamma\) lead to decreases in flexibility.

\(c,d\): Polynomial kernel parameters which when increased, increase the flexibility of the
decision boundary, and when small, decrease the flexibility of the decision boundary

When performing the grid search, a grid with a coarse granularity is selected with two benefits:
It reduces training time required and it prevents parameters being tuned too exactly to the
training data. A commonly used strategy for selecting \(\mathcal{C}\) is to search a logarithmic
grid, e.g.
\(\mathcal{C} \in \{10^{-N},10^{-N+1},\dotsc,10^{N}\}\)~\cite{kernel_methods_pattern_analysis,elem_stat_learn}.
We selected a smaller set of this range to search, depending on the kernel. For the linear
kernel \(\mathcal{C}\) was varied through \(\frac{10^N}{\mathit{D}_{mean}}\), \(N \in \{0,\dotsc,4\}\),
where \(\mathit{D}_{mean}\) is the mean Euclidean distance between training points from different classes.
For the polynomial and \gls{rbf} kernels, \(\mathcal{C}\) was varied over \(\{10^{0},\dotsc,10^{4}\}\).

The \gls{rbf} kernel parameter \(\gamma\) was selected such that a learned decision boundary is neither
too flexible nor insufficiently flexible. An estimate for the center of a search region is given as

\begin{figure}
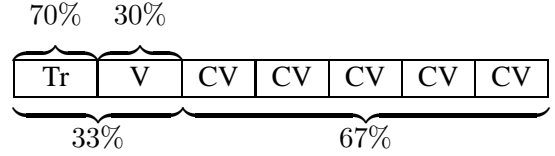

  \centering
  \begin{tabular}{ |c|c|c|c|c|c|c| }
  \multicolumn{1}{c}{\(70\%\)} & \multicolumn{1}{c}{\(30\%\)} & \omit & \omit & \omit & \omit & \omit \\
  \omit\mathstrut\downbracefill & \omit\mathstrut\downbracefill & \omit & \omit & \omit & \omit & \omit \\
  \hline
  Tr & V & CV & CV & CV & CV & CV \\
  \hline
  \omit\span\omit\mathstrut\upbracefill & \omit\span\omit\span\omit\span\omit\span\omit\mathstrut\upbracefill \\
  \multicolumn{2}{c}{\(33\%\)} & \multicolumn{5}{c}{\(67\%\)} \\
  \end{tabular}
  \caption{The data is partitioned into three groups; training, validation, and CV folds. After user selected
parameters are estimated with sets Tr and V, these sets are discarded, and 5-fold cross validation is performed
with the sets labelled CV}
  \label{fig:data_part}
\end{figure}

\begin{equation}
\label{equ:gamma_start}
\gamma_{start} = -\log_{10}\mathit{D}_{mean}~.
\end{equation}
Then, with center point specified by \eqref{equ:gamma_start}, the search region is selected as

\begin{equation}
\label{equ:gamma_search}
\gamma_{search} = 10^{N}, N \in [\gamma_{start}-2,\gamma_{start}+2]~.
\end{equation}

The polynomial kernel parameters \(c\) and \(d\) were again chosen to enable sufficient flexibility.
For this purpose, we chose \(d~\in~[2,6]\), allowing increasing flexibility with polynomial degree. In
order to limit the magnitude of the response of the polynomial kernel, we used the following estimate
for the starting point of \(c\);
\begin{equation}
\label{equ:poly_c_start}
c_{start} = \frac{1} {\displaystyle\argmax_{k} {\lvert\lvert x_i \rvert\rvert^2}}~,
\end{equation}
where \(x_i\) are the training data points. Then, to allow optimisation, a logarithmic range around \(c_{start}\) was considered,
\begin{equation}
\label{equ:poly_c_search}
c_{search} = c_{start}\cdot2^{2k}, k \in [-3,\dotsc,3]~.
\end{equation}

In order to ensure we have an good estimate of generalisation accuracy with limited training data,
we utilise data hold out, and cross validation. For that purpose, the data was partitioned
according to the scheme shown in \figurename~\ref{fig:data_part}.
For estimating user selected parameters, (e.g. the \gls{svm} cost parameter
\(\mathcal{C}\), \gls{rbf} parameter \(\gamma\), polynomial kernel parameters \(c\) and \(d\)),
one third of the
data is "held out" to be used for this estimation. Then, to estimate the generalisation accuracy, the
remaining data is split into equally sized sets. One of each of the sets is held out, while the remainder
is used for training a classifier (and estimating \acrlongpl{princomp}, when relevant). The held out set is
then used as a test set for recording classification accuracy. This is repeated with each of the sets used
once as a test set. Then, the mean accuracy and standard error across all test sets can be computed. 
When
comparing different classifiers, only the cross validation partitions are used for estimation
of generalisation accuracy, even if the method did not require parameter estimation, {\sl i.e.} 
\gls{qda} and \gls{lda} classifiers never see the data in the training and validation partitions.

\subsection{Multiclass classification using \gls{svm}s}
\label{subsec:svc_multiclass}

For multiclass discrimination, binary \gls{svm} classifiers are combined via predefined error-correcting output code
methods~\cite{multiclass_learning_problems_ecoc,multiclass_binary_approach_margin_classifiers}. To summarize the procedure briefly,
\(N\) binary classifiers are trained to distinguish between \(M\) classes using a coding matrix \(\textbf{W}_{M \times N}\),
with elements \(w_{mn}\in \{0,1,-1\}\). Classifier \(n\) is trained only on data of classes \(m\) for which \(w_{mn}\neq 0\),
with \(\mathrm{sgn}(w_{mn})\) as the class label. Then, the class assignment rule is given by

\begin{equation}
\label{equ:ecoc_prediction}
C(\x) = \argmin_m \sum_{n=1}^{N}\chi(w_{mn}f_{n}(\x))~,
\end{equation}
where \(f_{n}(\x)\) is the output of the \(n^\textrm{th}\) classifier and \(\chi\) is some loss function. 

The error-correcting capability of a code is commensurate with the minimum Hamming distance between the rows
of a coding matrix; if this minimum distance is \(\delta\), then the multiclass classifier will be able to
correct any \(\lfloor\frac{\delta - 1}{2}\rfloor\) errors~\cite{multiclass_learning_problems_ecoc}. For the
three class problem, \(\{SR,VT,VF\}\), we consider all \textit{one-vs-one} (pairwise) and all
\textit{one-vs-all} binary classifiers, which makes a total of six binary classifiers. Note that in the
case of three classes, this exhausts all possible binary classifiers. The corresponding coding matrix in this case thus has
the form

\begin{equation}
W =\left[
\begin{array}{rrrrrr}
	1  & 1  & -1 & 1  & 1  & 0\\
	1  & -1 & 1  & 0  & -1 & 1\\
	-1 & 1  & 1  & -1 & 0  & -1
\end{array}\right]~.
\end{equation}

A number of loss functions were compared, including hinge: \(\chi(z)=\max(1-z,0)\), Hamming: \(\chi(z)=[1-{\rm sgn}(z)]/2\),
exponential: \(\chi(z)=e^{-z},\) and linear: \(\chi(z)=-z\). The hinge loss function performed best and therefore is the only
loss function for which results are reported in Section \ref{subsec:svc_results}. Even so, the differences of decoding with
varying loss functions only changed by  few percent, with exponential loss typically performing the worst, and linear loss
performing almost as well as hinge loss.

\subsection{Classifier ensembles}
\label{subsec:ensembles}

Ensembles or committees of classifiers can often be combined to  improve classification accuracy~\cite{elem_stat_learn}. 
For that purpose we considered ensembles of decision values of binary \gls{svm} classifiers applied to \gls{ecg} segments of a given length,
taken with incremental shifts with respect to each other,
e.g. \(2\)~s segments taken over \(4\)~s intervals with \(0.5\)~s shifts to form ensembles of \(5\) decision values. 
The mean of decision values were then used as the corresponding values of \(f_n(\x)\) in the loss decoding given by (\ref{equ:ecoc_prediction}).
We also considered combining individual decision values within ensembles using median, majority vote, and maximum aggregation functions, but mean gave best results.

\section{Experimental Procedure and Results}
\label{sec:experiments}

\subsection{Data and preprocessing}
\label{subsec:data}

\subsubsection{Data sets}
Data were taken from Physiobank~\cite{physionet_generic}, which maintains a large online
repository of various physiological signals, including \gls{ecg} signals. The databases used
from Physiobank were the \gls{edb}~\cite{physionet_EDB}, the \gls{cudb}~\cite{physionet_CU}, the
\gls{mitdb}~\cite{physionet_BIH}, the \gls{vfdb}~\cite{physionet_VFDB}. Only data that were
explicitly labelled as \gls{srhythm}, \gls{v-tach} or \gls{v-fib} were used. Due to the fact that
the databases \gls{edb} and \gls{mitdb} do not contain many realisations of \gls{v-tach} or \gls{v-fib},
additional realisations of \gls{v-tach} and \gls{v-fib} are taken from \gls{vfdb}, and further \gls{v-fib}
realisations from \gls{cudb}. Neither of these databases contain annotations that have been audited thoroughly,
so they were only used to augment examples of ventricular arrhythmias, which are few in the other databases.

\subsubsection{Preprocessing}
In all these databases, \(250\)~Hz sampling rate is used, apart from \gls{mitdb} where signals are sampled at
\(360\)~Hz. It is considered that most of the relevant information is contained in the \(40\)~Hz baseband
\cite{arrhythmia_detection_dtw} and that preprocessing with a \(30\)~Hz low pass filter does not affect experimental results
\cite{wavelet_features_for_characterising_arrhythmia,vf_detection_emd,detecting_vf_time_delay_methods,
reliability_vf_detection_for_aed}. However, based on visual inspection of low-pass filtered data it was
decided that \(30\)~Hz cut-off frequency was too low, so \(49\)~Hz low-pass filtering was used, followed by
downsampling to \(100\)~Hz. In addition to this, a \(0.5\)~Hz high pass filter was applied to remove
wandering baseline~\cite{arrhythmia_detection_dtw}.
As in our previous study~\cite{eusipcopaper}, all \gls{ecg} records were normalised so
that the squared sum of each record is equal to the number of samples in the record,
thus making the variance of individual time samples equal to \(1\).

\subsubsection{Data balancing}
Using rhythm annotations, either \gls{srhythm}, \gls{v-tach}, or \gls{v-fib} were extracted from these databases.
The continuous sequences of the rhythms were then segmented according to window length for the experiment, and
shuffled randomly. Since there were many more examples of \gls{srhythm}, this class was randomly sub-sampled to
approximately the same amount as \gls{v-tach} and \gls{v-fib}, each of which had a total duration of approximately \(6000\)~s.
For several reasons, balanced sets of the three classes were used. First, it is hard to understand what the relevant class
priors should be, since it is highly context dependent, and also very hard to estimate. For example, in an \gls{icd},
which is always on, the presence of normal rhythm is much higher than arrhythmia, but exactly how much higher is hard
to quantify (e.g it depends on the specific patient's predisposition to an arrhythmia, underlying causes, etc). On the
other hand, an \gls{aed} is meant for use in emergency situations, {\sl e.g.} when a patient is unconscious or not breathing,
which makes the relative exposure of such a device to \gls{ecg} episodes of normal rhythm and arrhythmia considerably
different, and could be uniform or even skewed towards arrhythmia. Secondly, the \gls{svm} classification framework,
which is used in this study, does not support notion of priors in the formulation. Finally, balanced sets prevent
falsely high classification accuracy obtained by capitalising on high number of realisations of one of the classes.

\subsection{\gls{svm} classification results and comparison with \gls{qda}}
\label{subsec:svc_results}

\begin{figure}
  \centering
  \mbox{\subfloat[]{\label{subfig:svm_srvt_vf} \includegraphics[height=6cm]{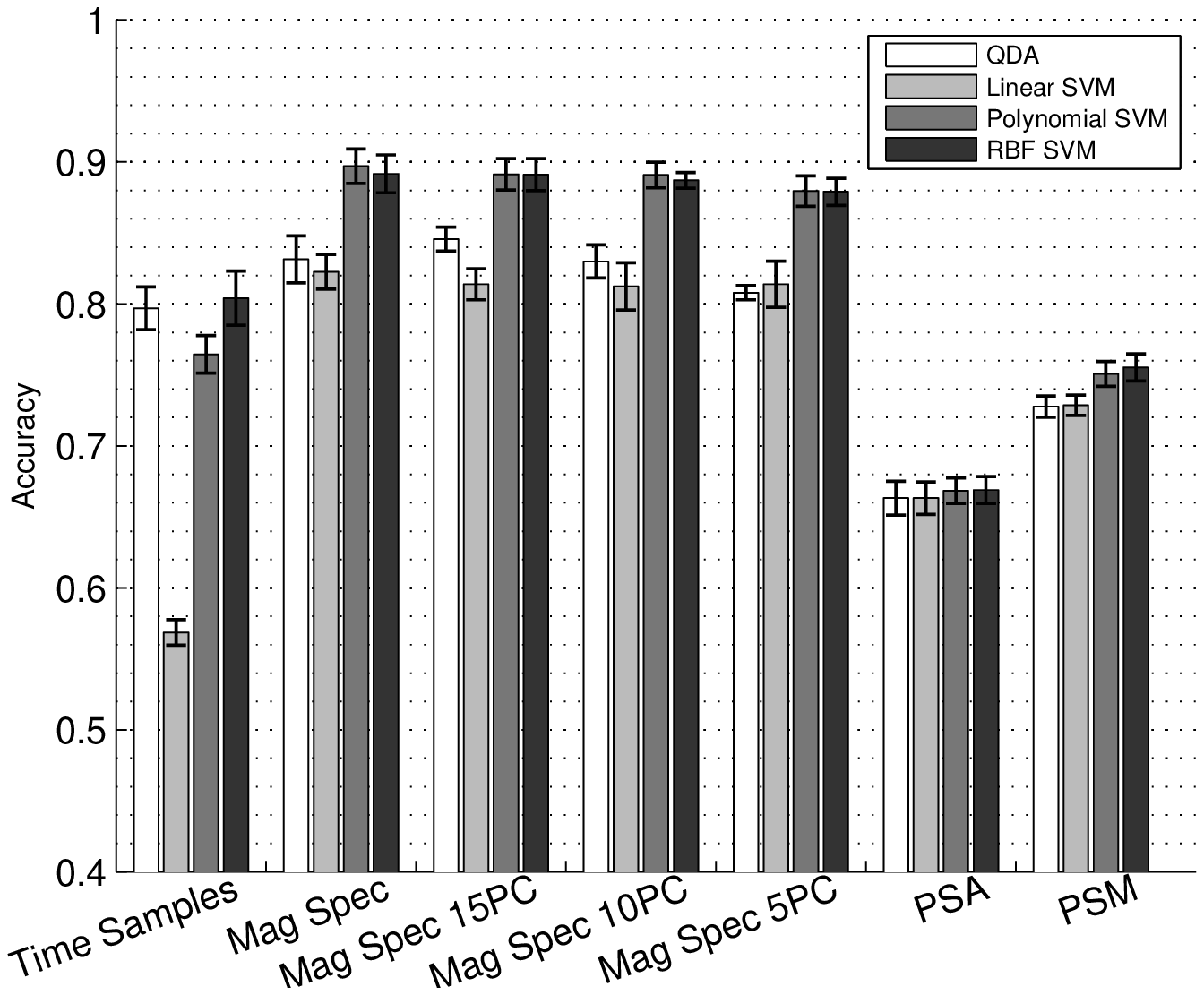}}}
  \mbox{\subfloat[]{\label{subfig:svm_vt_vf} \includegraphics[height=6cm]{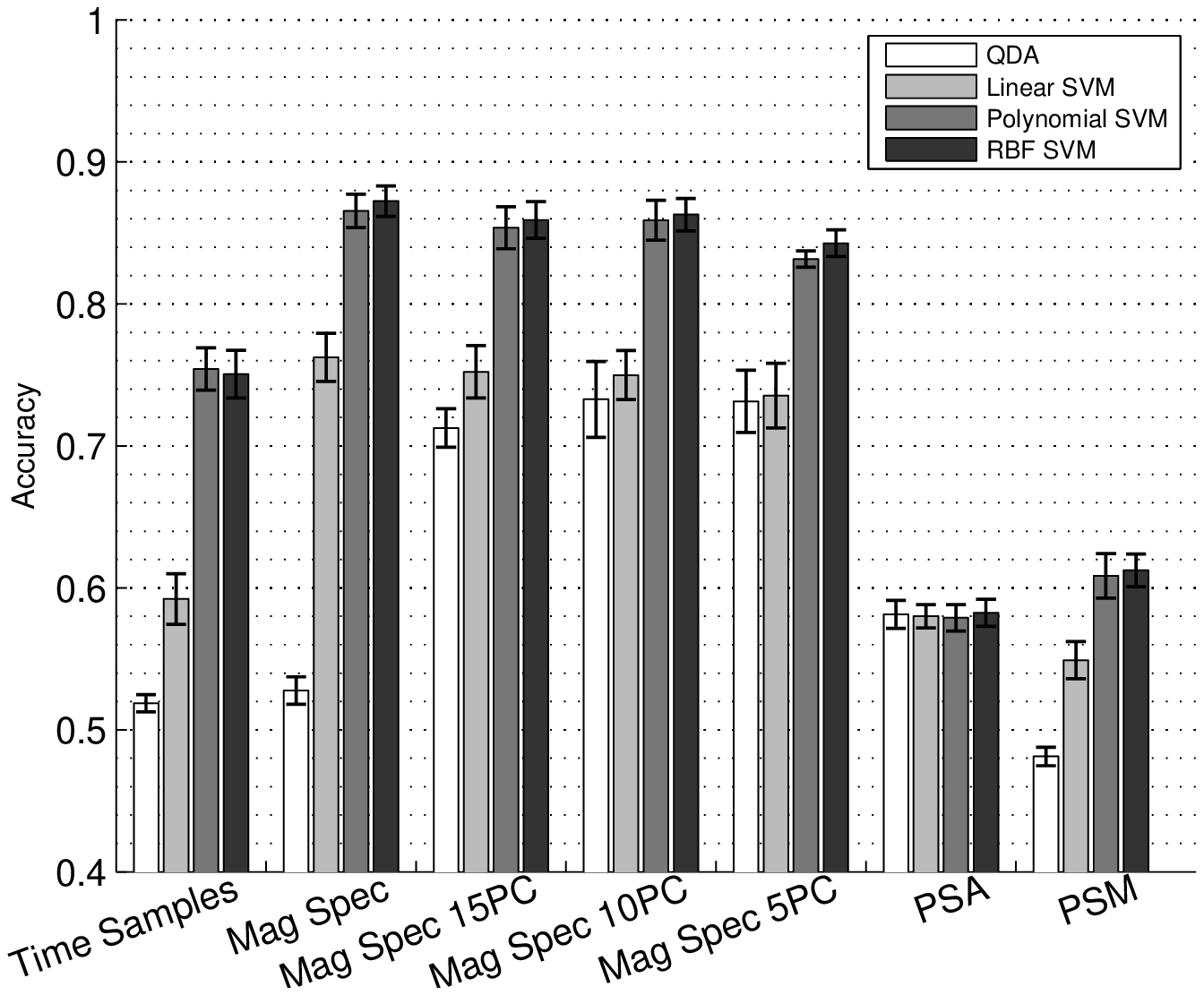}}}
  \mbox{\subfloat[]{\label{subfig:svm_3way} \includegraphics[height=6cm]{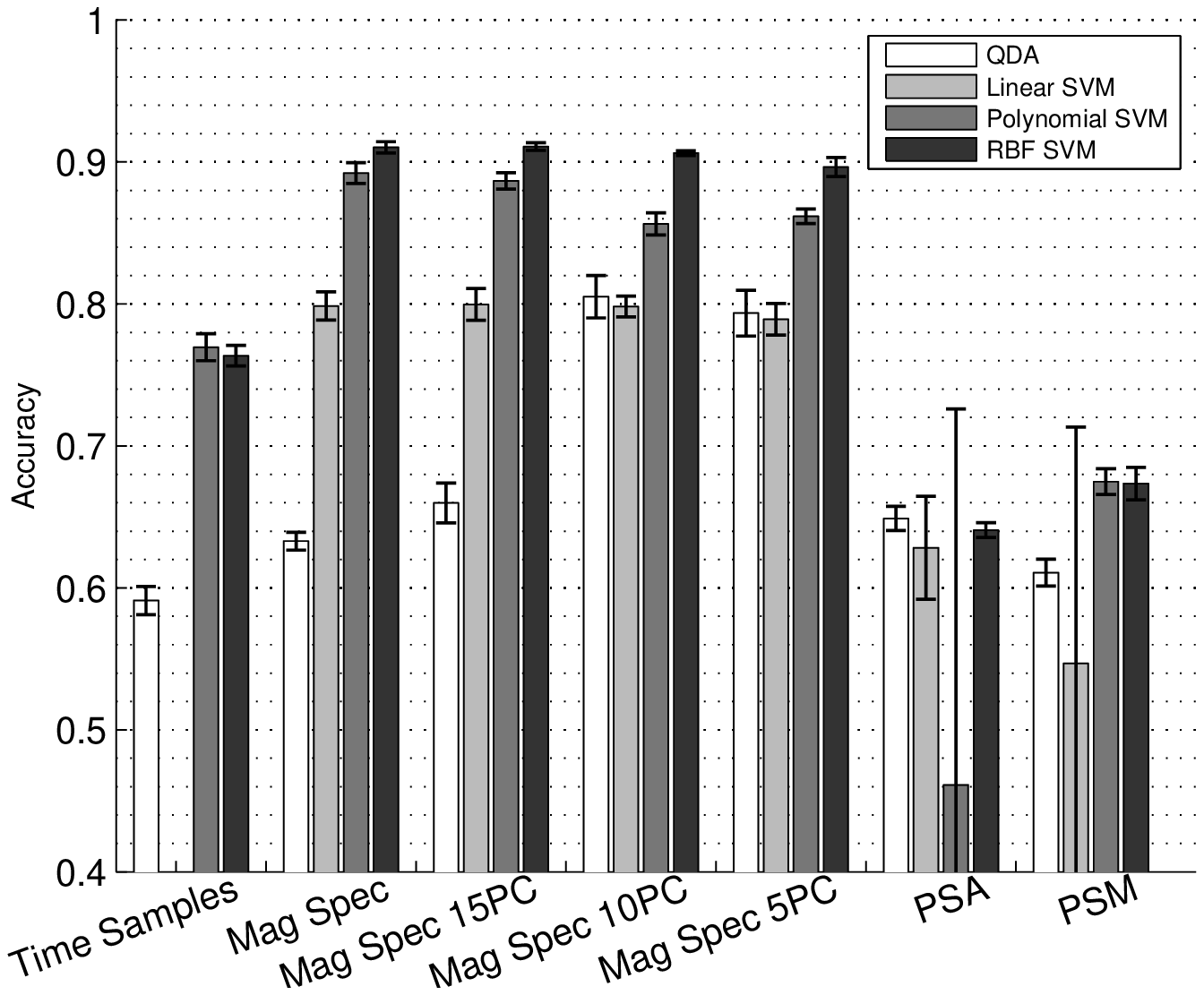}}}
  \caption{Average cross-validated classification accuracies with errorbars for 2~second observation length in the time, magnitude spectra, projections of magnitude spectra onto \acl{princomp} spaces, \gls{psa} and \gls{psm} representation spaces. Classification is performed with linear, \gls{rbf} and polynomial \glspl{svm} and compared with \gls{qda}. \protect\subref{subfig:svm_srvt_vf} non-\gls{v-fib} versus \gls{v-fib}. \protect\subref{subfig:svm_vt_vf} \gls{v-tach} versus \gls{v-fib}. \protect\subref{subfig:svm_3way} Three way classification between \gls{srhythm}, \gls{v-tach} and \gls{v-fib}}
  \label{fig:svm_classification}
\end{figure}

\begin{figure}
  \centering
  \includegraphics[height=6cm]{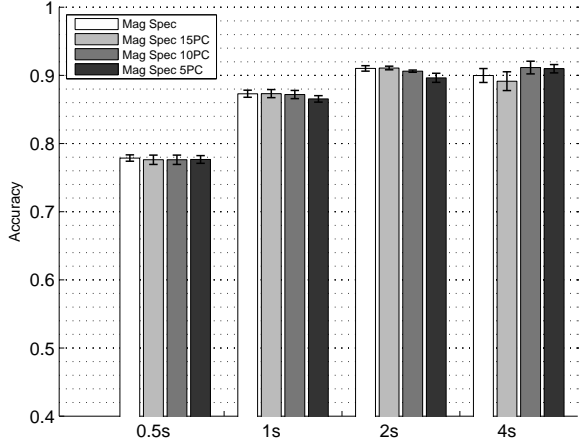}
  \caption{Average cross-validated accuracies with errorbars for classification between \gls{srhythm}, \gls{v-tach} and \gls{v-fib} using \(0.5\), \(1\), \(2\) and \(4\) second observation lengths in the magnitude spectra and projections of magnitude spectra onto \acl{princomp} spaces. Classification is performed with \gls{rbf} kernel \gls{svm}}
  \label{fig:rsvm_varying_windows}
\end{figure}

\figurename~\ref{subfig:svm_srvt_vf}, \figurename~\ref{subfig:svm_vt_vf} and \figurename~\ref{subfig:svm_3way} show results of classification
of non-\gls{v-fib} vs \gls{v-fib}, \gls{v-tach} vs \gls{v-fib}, and \gls{srhythm} vs \gls{v-tach} vs \gls{v-fib}, respectively in the time, magnitude spectra,
reduced magnitude spectra, \gls{psa}, and \gls{psm} representation spaces. This classification experiment was performed using \(2\)~s observation windows, as
our preliminary work showed that there is minor benefit in using longer segments~\cite{eusipcopaper}.
To assess benefits of more flexible non-linear class boundaries offered by the \gls{svm} framework, we considered \gls{svm} classification using polynomial and \gls{rbf} kernels in comparison with linear kernels and \gls{qda}.

First it can be observed that classification using magnitude spectra or their reduced versions achieves much higher accuracy than classification using time-domain waveforms or \gls{psa}/\gls{psm} representations. This is in agreement with results of our preliminary analysis which used \gls{lda} and \gls{qda} classification~\cite{eusipcopaper}. Hence, time-domain \gls{ecg} waveforms, \gls{psa} and \gls{psm} were not considered any further.
In most cases at least 5\% improvement in classification accuracy over \gls{qda} or linear \gls{svm}s is achieved by use of \glspl{svm} with non-linear kernel functions (\figurename~\ref{fig:svm_classification}).

Corresponding sensitivity values (not included in the paper) revealed a bias in classification, usually towards \gls{srhythm}.
The average sensitivities for \gls{v-tach} and \gls{v-fib} did not exceed 90\%, despite achieving a best accuracy of 91\% with a very small standard error. The bias was more pronounced with certain classifiers (\gls{qda}) and representation spaces (time-domain \gls{ecg}, phase space representations).

\begin{table}
\centering
\footnotesize
\caption{Average sensitivities for \gls{srhythm}, \gls{v-tach}, \gls{v-fib} for classification of various observation lengths using \gls{rbf} \gls{svm} in magnitude spectra, and reduced magnitude spectra representation spaces. Each row corresponds with a bar from \figurename~\ref{fig:rsvm_varying_windows}. }
\label{tab:rsvm_varying_windows}
\begin{filecontents*}{data/rsvm_varying_windows_sns.csv}
"0.5 seconds","Magnitude Spectra",91.8,69.8,72.0
,"15PCs",91.4,70.2,71.2
,"10PCs",91.4,70.2,71.2
,"5PCs",92.0,69.9,71.1
"1 second","Magnitude Spectra",95.3,80.4,86.2
,"15PCs",95.3,80.4,86.3
,"10PCs",95.4,80.1,86.1
,"5PCs",95.4,79.4,84.9
"2 seconds","Magnitude Spectra",97.9,86.0,89.2
,"15PCs",98.2,86.2,89.0
,"10PCs",97.8,85.7,88.3
,"5PCs",97.9,83.5,87.6
"4 seconds","Magnitude Spectra",97.0,88.0,84.9
,"15PCs",96.1,85.9,85.4
,"10PCs",96.9,88.8,87.7
,"5PCs",96.9,87.6,88.4
\end{filecontents*}
\pgfplotstabletypeset[
    begin table=\begin{tabular},
    header=false,
    ignore chars={"},
    every head row/.style={before row=\toprule, after row=\midrule},
    every last row/.style={after row=\bottomrule},
    every nth row={4}{before row=\midrule},
    string type,
    columns/0/.style={column name=\textbf{Window Size}, column type={l}},
    columns/1/.style={column name={\begin{minipage}{2cm}\centering\textbf{Representation space}\end{minipage}}, column type={l}},
    columns/2/.style={column name={\begin{minipage}{0.7cm}\vspace{0.1cm}\centering\textbf{\gls{srhythm} sens. (\%)}\end{minipage}}, column type={c}},
    columns/3/.style={column name={\begin{minipage}{0.7cm}\vspace{0.1cm}\centering\textbf{\gls{v-tach} sens. (\%)}\end{minipage}}, column type={c}},
    columns/4/.style={column name={\begin{minipage}{0.7cm}\vspace{0.1cm}\centering\textbf{\gls{v-fib} sens. (\%)}\end{minipage}}, column type={c@{}}},
    end table=\end{tabular}
    ]{data/rsvm_varying_windows_sns.csv}
\end{table}

Having established that the best performing representations were the magnitude spectra and their subspaces, and 
that \gls{svm}s with \gls{rbf} kernels achieved highest accuracy, the issue of window length was revisited. 
\figurename~\ref{fig:rsvm_varying_windows} shows results of 
\gls{srhythm} versus \gls{v-tach} versus \gls{v-fib} classification in magnitude spectra representation spaces using \gls{svm}s with \gls{rbf} kernels for observation lengths of \(0.5\)~s, \(1\)~s, \(2\)~s, and \(4\)~s. The corresponding sensitivities of individual classes are shown in 
\tablename~\ref{tab:rsvm_varying_windows}. The overall accuracy and the
sensitivities improved with increasing observation lengths up to \(2\)~s when a performance plateau is attained,
in agreement with our preliminary study in which \gls{lda}, and \gls{qda} were used~\cite{eusipcopaper}. 
The fact that increasing the segment length from \(2\)~s to \(4\)~s did not affect classification accuracy inspired forming ensembles of classifiers 
acting on segments taken with incremental shifts over a larger observation window. 

\begin{figure}
  \centering
  \includegraphics[height=6cm]{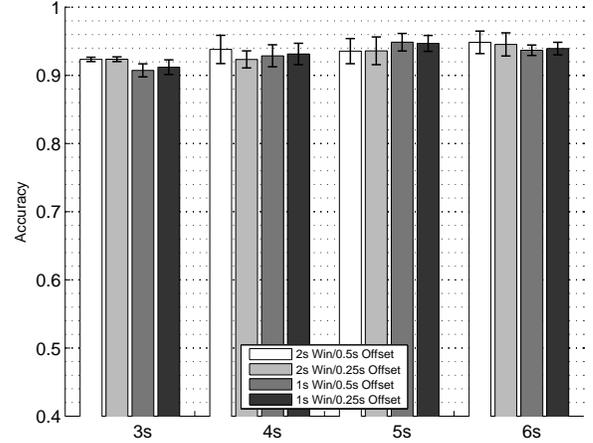}
  \caption{Average cross-validated accuracies with errorbars for classification between \gls{srhythm}, \gls{v-tach} and \gls{v-fib} using \(3\), \(4\), \(5\) and \(6\) second observation lengths in the magnitude spectra representation space with \gls{rbf} \gls{svm} classifiers. Ensembles were obtained by using \(2\) second or \(1\) second windows, and \(0.5\) second or \(0.25\) second offsets.}
  \label{fig:window_ensembles}
\end{figure}

\begin{table}
\centering
\footnotesize
\caption{Average sensitivity values for each class, \gls{srhythm}, \gls{v-tach}, \gls{v-fib} shown for different window lengths, with different options for forming ensembles from the given window length. Each row corresponds with a bar from \figurename~\ref{fig:window_ensembles}. The representation space is magnitude spectra, and the classifier is \gls{rbf} \gls{svm}. }
\label{tab:window_ensembles}
\begin{filecontents*}{data/window_ensembles_sns.csv}
"3 seconds","Window 2s / Offset 0.5s",97.7,88.8,90.5
,"Window 2s / Offset 0.25s",97.9,88.9,90.3
,"Window 1s / Offset 0.5s",97.4,87.2,87.7
,"Window 1s / Offset 0.25s",98.3,87.3,88.0
"4 seconds","Window 2s / Offset 0.5s",97.2,92.8,91.4
,"Window 2s / Offset 0.25s",98.7,88.4,89.9
,"Window 1s / Offset 0.5s",99.0,89.5,90.1
,"Window 1s / Offset 0.25s",99.1,89.9,90.4
"5 seconds","Window 2s / Offset 0.5s",98.1,90.5,92.1
,"Window 2s / Offset 0.25s",98.0,90.7,92.2
,"Window 1s / Offset 0.5s",98.2,93.2,93.2
,"Window 1s / Offset 0.25s",98.2,93.2,92.7
"6 seconds","Window 2s / Offset 0.5s",98.6,92.7,93.3
,"Window 2s / Offset 0.25s",98.3,93.0,92.4
,"Window 1s / Offset 0.5s",97.1,92.4,91.6
,"Window 1s / Offset 0.25s",97.1,93.1,91.6
\end{filecontents*}
\pgfplotstabletypeset[
    begin table=\begin{tabular},
    header=false,
    ignore chars={"},
    every head row/.style={before row=\toprule, after row=\midrule},
    every last row/.style={after row=\bottomrule},
    every nth row={4}{before row=\midrule},
    string type,
    columns/0/.style={column name=\textbf{Observation}, column type={l}},
    columns/1/.style={column name=\textbf{Ensemble Options}, column type={l}},
    columns/2/.style={column name={\begin{minipage}{0.7cm}\vspace{0.1cm}\centering\textbf{\gls{srhythm} sens. (\%)}\end{minipage}}, column type={c}},
    columns/3/.style={column name={\begin{minipage}{0.7cm}\vspace{0.1cm}\centering\textbf{\gls{v-tach} sens. (\%)}\end{minipage}}, column type={c}},
    columns/4/.style={column name={\begin{minipage}{0.7cm}\vspace{0.1cm}\centering\textbf{\gls{v-fib} sens. (\%)}\end{minipage}}, column type={c@{}}},
    end table=\end{tabular}
    ]{data/window_ensembles_sns.csv}
\end{table}

\begin{figure}
  \centering
  \mbox{\subfloat[]{\label{subfig:vt_gold_standard} \includegraphics[height=6cm]{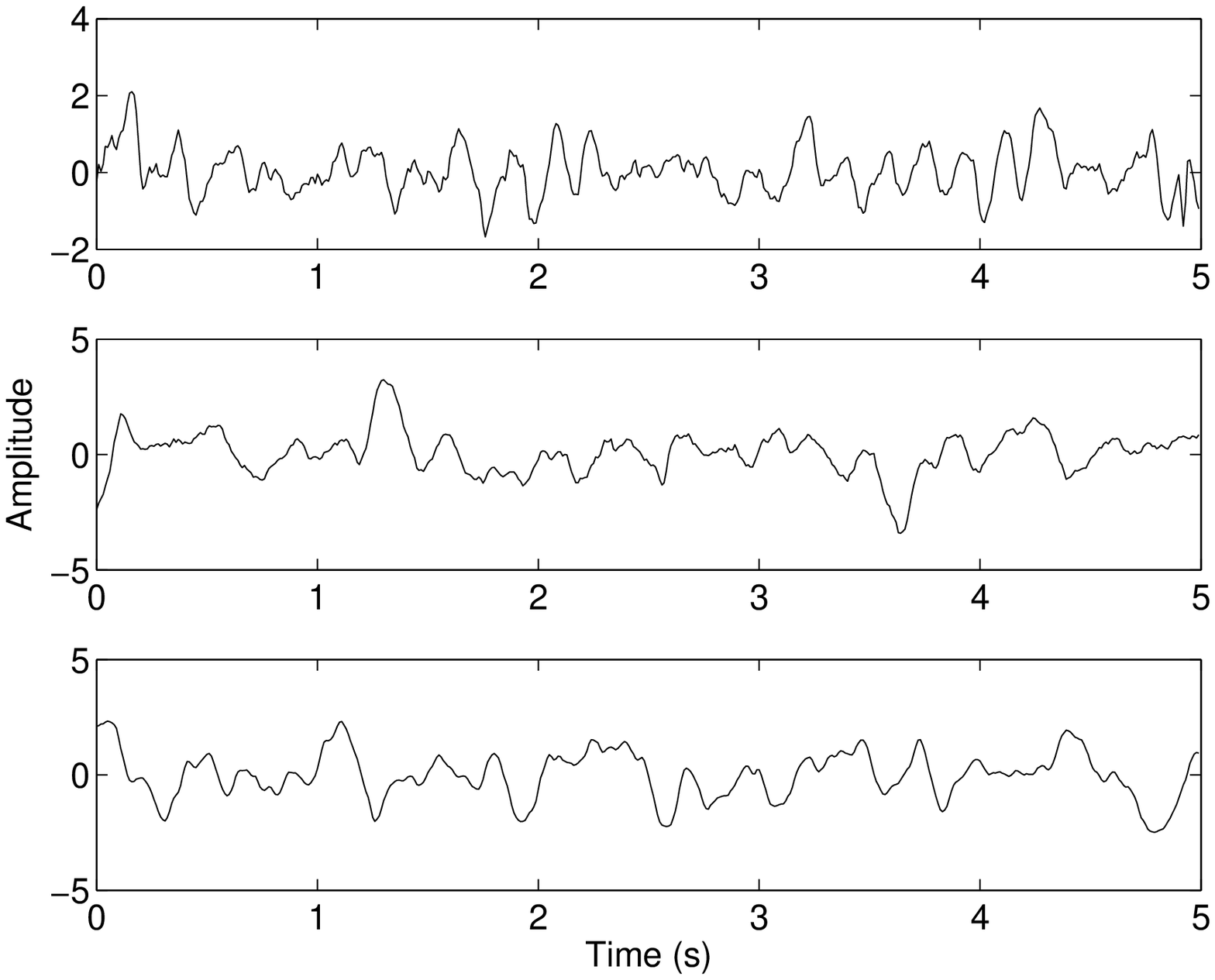}}}
  \mbox{\subfloat[]{\label{subfig:vf_gold_standard} \includegraphics[height=6cm]{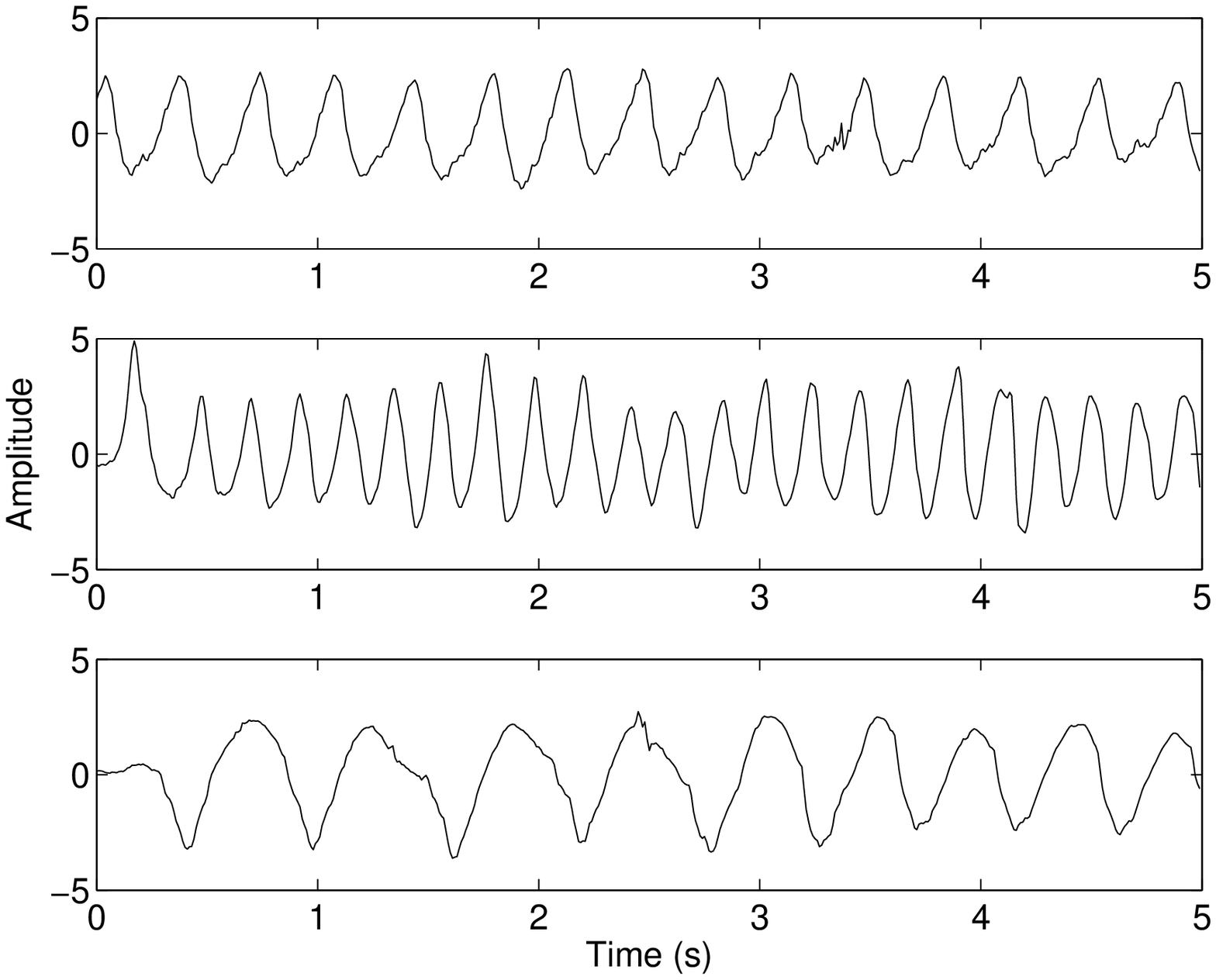}}}
  \caption{Examples of ``misclassifications'' by the best classifier from \figurename~\ref{fig:window_ensembles}. \protect\subref{subfig:vt_gold_standard} rhythms labeled as \gls{v-tach}, that were classified as \gls{v-fib}. \protect\subref{subfig:vf_gold_standard} rhythms labeled as \gls{v-fib}, that were classified as \gls{v-tach}.}
  \label{fig:misclassified}
\end{figure}

\figurename~\ref{fig:window_ensembles} shows results of classification between \gls{srhythm}, \gls{v-tach} and \gls{v-fib} using \gls{svm} decision
values averaged
over an ensemble of \gls{ecg} segments, extracted from longer windows, as specified in Section~\ref{subsec:ensembles}.
These shorter segments were transformed to the best performing representation space, magnitude spectra, and classified using \gls{svm}s with 
\gls{rbf} kernels. For this purpose \(1\)~s and \(2\)~s segments for classification were considered, taken with \(0.25\)~s and \(0.5\)~s shifts from 
\(3\)~s, \(4\)~s, \(5\)~s and \(6\)~s windows. For ensembles formed over \(5\)~s or longer, all variants performed with 90\% or greater sensitivity for all classes (\tablename~\ref{tab:window_ensembles}). The best performing ensemble was obtained with \(1\)~s segments, shifted by \(0.5\)~s, extracted from larger windows of \(5\)~s. 
Average sensitivities obtained with these ensembles were above 93\% for all classes. For these parameters, we show in \figurename~\ref{fig:misclassified} some examples of misclassified data. \figurename~\ref{subfig:vt_gold_standard} shows three examples of rhythms labelled as \gls{v-tach} by the databases, and classified as \gls{v-fib} by the classifier; \figurename~\ref{subfig:vf_gold_standard} shows three examples of rhythms labelled as \gls{v-fib} by the databases, and classified as \gls{v-tach} by the classifier. In our opinion, these examples are classified correctly by the algorithm and mislabelled in the databases. Thus, the databases themselves impose a limitation on the best achievable accuracy, but also limit what is learnable from the data, particularly if considerable amounts of data are mislabelled.

\section{Conclusion}
\label{sec:conclusion}

A systematic study of \gls{ecg} rhythm classification was presented, to develop methods for accurate classification
between \gls{srhythm}, \gls{v-tach} and \gls{v-fib}. Classification was performed in the space of \gls{ecg} waveforms, their magnitude spectra, and lower-dimensional approximations of magnitude spectra using projections onto their \acl{princomp} spaces. All representation spaces are of much higher dimension than considered in previous studies.
We found that  classification accuracy improves with observation length up to about \(2\)~s, where 
it reaches a plateau, which is significantly shorter than used in previous
studies~\cite{reliability_vf_detection_for_aed,detecting_vf_time_delay_methods,detection_arrhythmia_using_crosscorrelation}.
Since \gls{v-fib} can be highly transient in some animal models this may assist translational research.

Among considered representation spaces and classification methods, magnitude spectra and \gls{svm} classifiers
with \gls{rbf} kernels achieved highest accuracy. In combination they achieved classification accuracy of 91\% and sensitivity of at least 86\% for each class
using only 2~s segments of \gls{ecg} signals. However, ensembles of \gls{svm} classifiers applied to \(1\)~s segments  over a \(5\)~s intervals reduced classifier bias and achieved 94\% classification accuracy, with sensitivity for each class at 93\% or higher.

\ifCLASSOPTIONcaptionsoff
  \newpage
\fi



\bibliographystyle{IEEEtran}
\bibliography{bibs/classification_papers,bibs/physiology,bibs/signal_processing,bibs/misc}
%

\end{document}